**Resources for Supporting Students with and without Disabilities in Your Physics Courses**

Amanda Lannan[1], Erin Scanlon[2], & Jacquelyn J. Chini[2]

[1] College of Education and Human Performance, University of Central Florida
[2] Physics Department, University of Central Florida

1. Introduction

Students with disabilities are enrolling in postsecondary education in increasing numbers[1] and in science, technology, engineering, and mathematics (STEM) at steady rates since the early 1990s.[2] Specifically, in 2014, the National Center on Science and Engineering Statistics (NCSES) found that 10.5% of students enrolled in science and engineering degree programs identified with a disability.[3] However, postsecondary faculty have been shown to be unprepared to support students with disabilities in their classes[4] and popular, research-based introductory physics curricula do not adequately plan for variations in learners' needs, abilities, and interests.[5] The purpose of this paper is to provide resources that instructors can use in their classes to promote accessibility and support all learners. In this paper we: 1) provide a brief review of the literature related to supporting students with disabilities in the context of physics; 2) describe a design framework intended to encourage development of curricula that support all learners; and 3) provide a list of resources that physics instructors can use to increase support for students with disabilities.

2. Supporting Students with Disabilities in Postsecondary STEM

Most of the literature about supporting students with disabilities in physics target students with visual impairments (59% of articles we identified; e.g., [6-10]), physical disabilities (23%; e.g., [11-15]), and hearing impairments (18%; e.g., [16-19]). We did not identify any articles related to supporting students with affective, emotional/behavioral, or mental health impairments. Most articles described new or adapted apparatus to support people with vision, hearing, or physical impairments.[20-22]

We found that the same topics were revisited by multiple papers across decades. For example, the following articles all discuss how to support students who are blind in physics courses.[23-28] Additionally, there is not much uptake of these articles in the literature as shown by low citation rates (i.e., the 22 articles we identified had average yearly citation rates of 0.64, with a minimum of 0 and a maximum of 1.22; total citations ranged from 0 to 47 per article, with an average of 8).

Postsecondary institutions that receive federal funding must comply with several laws that mandate equal access for qualified students with disabilities. The Individuals with Disabilities Education Act[29], Section 504 of the Rehabilitation Act of 1973[30], and Title II of the Americans with Disabilities Act of 1990[31] are antidiscrimination laws that protect individuals with disabilities who attend postsecondary institutions receiving federal funding. However, unlike K-12 settings, the responsibility is on postsecondary students to self-disclose their disability status to access supports. Coordination of accommodations and services are often facilitated by an Office for Students with Disabilities. Disability disclosure may make students vulnerable to stigmatization from their instructors and peers.[32] Similarly, James, Bustamante, Lamons, and Chini (2018) interviewed students with disabilities about their experiences in an introductory physics course and found the same trend.[33]

3. Universal Design for Learning (UDL)

One way to provide support for students with disabilities without requiring disclosure is to proactively make the learning environment accessible to and supportive of all learners from the beginning instead of reactively providing supports for individual students' needs. We suggest instructors use the Universal Design for Learning (UDL) framework during the development of curricular materials and implement UDL-aligned strategies in their courses. UDL is a framework for the development of curricular materials that are designed to be inherently accessible to and supportive of all learners. This framework is underpinned by the idea that there is no "average" student, which is supported by current neurological research.[34] All people inherently vary in their needs, abilities, and interests along a multi-dimensional spectrum,[35] so instructors can proactively design curricula with this variation in mind without knowing the specific impairments experienced by students in a given class.

The UDL framework is composed of three guidelines: 1) provide multiple means of representation (presenting information in multiple formats such as text and diagrams); 2) provide multiple means of action and expression (providing options in how students express their understanding); and 3) provide multiple means of engagement (provide options to support variation in students' motivations and interests). Each guideline is further described by three principles and a total of 30 smaller-grained checkpoints. Figure 1 shows a visual representation of the three guidelines and nine principles in the UDL 2.2 framework.[36]

Figure 1: Universal Design for Learning framework 2.2[36]

|  | Provide Multiple Means of Engagement | Provide Multiple Means of Representation | Provide Multiple Means of Action and Expression |
|---|---|---|---|
| Access – provide options for: | Recruiting interest | Perception | Physical action |
| Build – Provide options for: | Sustaining effort and persistence | Language and symbols | Expression and communication |
| Internalize – Provide Options for: | Self-regulation | Comprehension | Executive functions |
| Goal – Expert learners who are: | Purposeful and motivated | Resourceful and knowledgeable | Strategic and goal directed |

While UDL is often associated with students with disabilities, it is a useful framework to plan for all dimensions of learner variability[35] and promotes instructor emphasis on what students do well rather than their limitations.[37] For example, adding captions to instructional videos benefits both students with hearing impairments as well as students accessing the material while their children are sleeping or from a noisy team bus.

3.1 Designing Physics Courses with UDL in Mind

Instructors may feel overwhelmed by the task of developing a course that maintains high rigor while incorporating the flexibility of UDL. UDL is a framework grounded in the neuroscience of why (multiple means of engagement), what (multiple means of representation), and how (multiple means of action and expression) people learn.[38] Therefore, the first step to implementing Universal Design is to examine the why, what, and how of our teaching while looking for the barriers our students frequently encounter. Instructors should ask themselves: "Why should students care about this topic?"; "What do students find challenging about this topic?"; "How do students show their understanding of this topic?"

Next, instructors should identify or develop strategies to provide support and options. For example, an instructor could examine how she covers momentum in her introductory physics course by first thinking about the connections between momentum and her students' interests, major discipline, and future careers. After the instructor has identified these connections, she could add explicit mentions of these connections in class to provide one option for motivation. The following section describes resources that may assist physics instructors in the process of identifying and developing strategies.

4. Resources

In this section we describe several resources that may assist physics instructors in creating flexible options for how students access content, demonstrate their understanding, and engage with the course. These options include off-the-shelf products, assistive technologies, and high-tech and low-tech options. In some cases, technological resources can be leveraged to reduce or even eliminate barriers for some students with disabilities; in all cases, resources should be used in a way that supports all students in optimizing their learning. Access to a variety of tools, resources, and options provides ALL students the flexibility to interact with the course in a way that best supports their learning.

4.1 Alternative Access to Print Materials

Image and text-based information is a central component of physics courses. For many students, commercially available print or e-textbooks sufficiently support their learning. However, some students (e.g., those with visual impairments, learning disabilities, attention deficits, neurological disorders) may experience difficulty accessing information presented in traditional formats. Instructors can proactively accommodate students with such impairments and provide choice to all students by providing text and image-based information in alternative formats.

Reading systems (e.g., assistive technology that provides access to text-based materials) must include accessibility features, such as support for a screen reader, ability to change the visual presentation (e.g., customizing screen layout), read aloud feature, and text feature adjustments (e.g., enlarge text and adjust contrast). Common reading applications include: Vital Source,[39] Dolphin EasyReader,[40] and Voice Dream Reader.[41] These applications are available for Windows, Mac, iOS, and Android operating systems and via a browser with customizable options. Special library services, such as Book Share,[42] Learning Ally,[43] and the national library services,[44] can provide audio and braille textbook files to individuals with qualifying impairments.

Physics instructors must also consider how to provide alternative access to diagrams, figures, and graphs. Tactile representations can be created using 3D printers. For example, see 3D OPAL (3D Objects for Physics Accessible Learning) project which has plans for physics-specific objects.[45] Figure 1 shows an example of a 3D OPAL tactile representation of

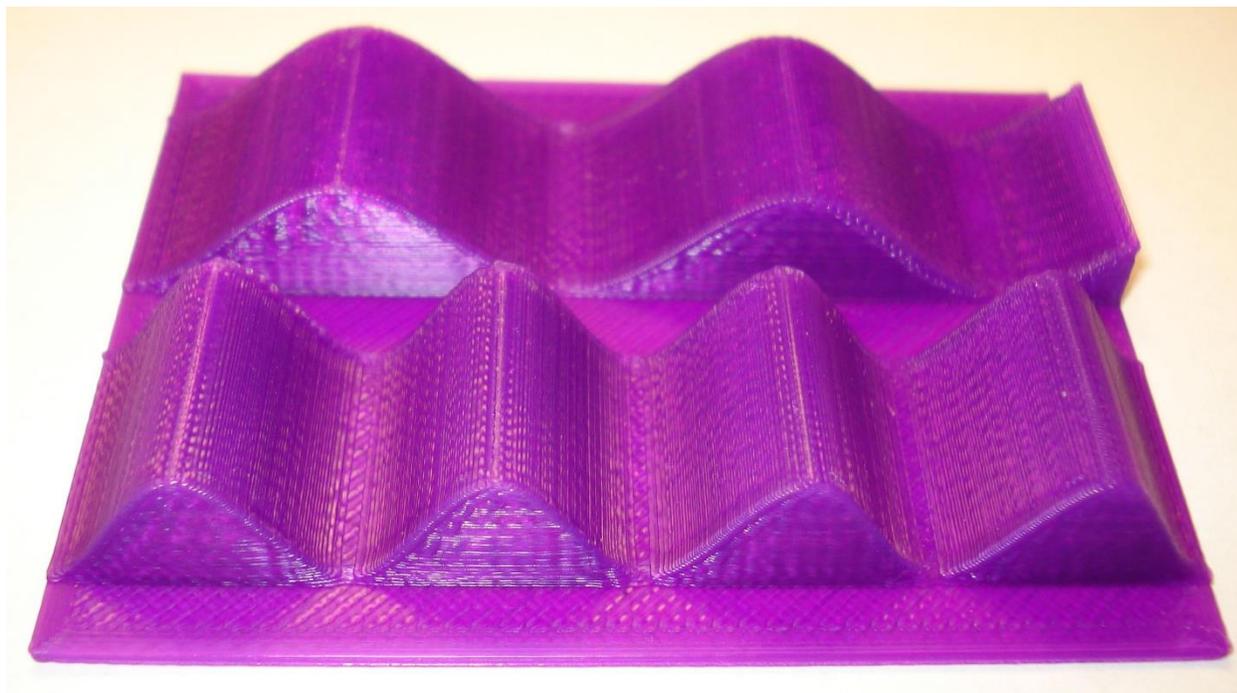

Figure 1: 3D OPAL tangible representation of standing waves with different wavelengths. (Figure used with permission)

The Sensational Blackboard[46] can instantly create tactile diagrams using only standard copy paper and a ballpoint pen. 3Doodler Start[47] is an inexpensive solution for creating 3D images in the air or on a flat surface. These commercially available products can create tactile representations on the spot during a class session whereas the 3D OPAL representations must be printed before class.

4.2 Virtual Labs

For students with dexterity or mobility impairments, virtual environments may offer increased independence over traditional laboratory experiences. For example, an introductory circuits lab experiment could be conducted with a PhET simulation (Physics Education Technology)[48] in addition to physical equipment, allowing students with dexterity impairments to create and test circuits. Many virtual labs are accessible outside of class, which allows students with difficulties maintaining attention throughout a long laboratory class period (e.g., students with ADHD, learning disabilities) to repeat the experiment at home. Virtual labs can be found at the PHET website,[49] via the University of Oregon,[50] or at the Teaching Commons.[51] It is important for instructors to consider the accessibility of the simulations they choose. For example, some PhET simulations incorporate accessibility features such as alternative input, simple descriptions via screen reader software, dynamic description, and sound and sonification.[52]

4.3 Accessible Data Analysis

Many physics courses involve calculator use, and traditional calculators may create barriers for some students with disabilities. For example, button manipulation can be a barrier for people with dexterity impairments and visual outputs are a barrier for people with vision impairments. Talking calculators, such as the ORION TI-34 Talking Scientific Calculator or the Audio Graphing Calculator for Windows, provide students access to calculations and graphing functions through spoken menus, scalable visual displays, keyboard navigation, sonification, and tactile output options.

Some instructors use computer-based programs (e.g., Microsoft Excel, Python, Matlab) for manipulating, analyzing and presenting large data sets. One way such data sets are made accessible is by conversion to tactile graphs; while this method can give students access to overall trends, it can be tedious and/or complicated to access more specific information. Alternatively, information can be represented via data sonification,[53] where each numeric value in a data column is connected to an audio frequency, allowing the user to listen to the trends in the raw data. Some programs, such as SAS Graphics Accelerator, have built-in data sonification features that generate alternative presentations of data visualizations including text descriptions, tabular data, and interactive sonification.

4.4 Physical Layout

Instructors should also consider whether the physical design of their instructional space is inclusive and welcoming. For example, all students, including those who use wheelchairs, should be able to move through the space easily and reach table surfaces, storage areas, and sinks. Instructors can solicit students' seating preferences to facilitate needs related to visual, auditory, or attention impairments. For example, one student may prefer to sit near the speaker if they use lip reading and another student may feel anxious when not seated near the exit. Cameras and projectors or ceiling-mounted mirrors (e.g., Sheldon Laboratory Systems Visual Aider Overhead Mirror[54]) can improve students' visual access to demonstrations, and microphones can be used to improve auditory access.

5. We Must Also Provide Training to Use Accessible Tools and Technologies

For new resources to support learners, instructors and students must have access to training about using the resource. Without training, new resources can become a barrier rather than a support. Thus, we advocate for curriculum developers to work with accessibility experts to provide recommendations for assistive technologies and training guides that describe how the curriculum incorporates the technology. For example, if a student is required to use an accessible calculator, training should be available that describes: how the calculator interfaces with the user, how the user interfaces with the calculator, and how to use the device effectively in the context of the course. We invite the *TPT* community to join this effort by contributing articles related to supporting students with a broad range of impairments and/or abilities in becoming expert physics learners.

6. Further Reading and Resources

- More information about the Universal Design for Learning framework can be found at the Center for Applied Special Technology (CAST)[55] and in the book *Academic Ableism: Disability and Higher Education*[56].
- For physics-specific recommendations, see *A Guide to Disability Good Practice for University Physics Departments*.[57]
- The CAST[55] website also catalogs examples of colleges and universities in varying phases of implementing UDL initiatives. The UDL ON CAMPUS website (http://udloncampus.cast.org/home) describes professional development options and resources, including syllabus design, environmental considerations, and technology solutions. For example, UDL-Universe (UDL-U) is a multi-campus initiative through the California State University System, offering comprehensive guidance for UDL course re-design. At Colorado State University, technical modules, student self-advocacy resources, and information about disabilities and accommodations, are available.